\RequirePackage[hyphens]{url}
\documentclass[transmag]{IEEEtran}

\usepackage{latexsym}
\usepackage{amsmath,amssymb,amsfonts}
\usepackage{mathtools}
\usepackage{algorithmic}
\usepackage{graphicx}
\usepackage{textcomp}
\usepackage{xcolor}
\usepackage{multicol}
\usepackage[USenglish]{babel} 
\usepackage[T1]{fontenc}
\usepackage[ansinew]{inputenc}
\usepackage[noadjust]{cite}
\usepackage{lmodern} 
\usepackage{graphicx} 
\usepackage{float}
\usepackage{bm}
\usepackage{bbm}
\usepackage{url}
\usepackage{hyperref}
\usepackage{booktabs} 
\usepackage{subcaption}
\usepackage{caption} 
\usepackage[font=small,labelfont=bf]{caption}
\usepackage{textcomp}
\usepackage{tabularx}
\usepackage{multirow}

\renewcommand{\baselinestretch}{0.97} 

\newcommand{\xstate}{\bm{x}}	
\newcommand{\Ostate}{{O}}
\newcommand{\Obstate}{\tilde{{O}}}
\newcommand{\trstatus}{\bm{tr_{\text{status}}}}
\newcommand{\ddim}{\bm{d}}

\newcommand{\fv}{\text{fov}}
\newcommand{\Los}{\text{LoS}}
\newcommand{\sco}{{sc}}
\newcommand{\isconf}{{cnf}}
\newcommand{\iscoast}{{cst}}
\newcommand{\Olistloc}{\mathcal{O}_{\text{}}}
\newcommand{\Olistsys}{\tilde{\mathcal{O}}}
\newcommand{\pp}{p_{\exists}}
\newcommand{\spp}{s_{\exists}}
\newcommand{\lvru}{l_{\text{\tiny vru}}}
\newcommand{\vvru}{v_{\text{\tiny vru}}}

\newcommand{\unk}{\{\exists, \nexists\}}	
\newcommand{\ptrust}[1]{p_{\text{trust}}^{#1}}	
\newcommand{\pex}[2]{p_{\text{ex}}^{#1}(#2)}	
\newcommand{\pfov}[2]{p_{\text{FoV}}^{#1}(#2)}	
\newcommand{\pocc}[2]{p_{\text{occ}}^{#1}(#2)}	
\newcommand{\pdm}[2]{p_{\text{dm}}^{#1}(#2)}	
\newcommand{\pval}[2]{p_{\text{val}}^{#1}(#2)}	
\newcommand{\pany}[1]{p_{\text{#1}}}	
\newcommand{\pfa}{p_{\text{fa}}}	
\newcommand{\pd}{p_{\text{d}}}	
\newcommand{\tMR}{\text{MR}}
\newcommand{\tUOR}{\text{UOR}}
\newcommand{\hist}{\text{hist}}
\newcommand{\dimvel}{\text{dim-vel}}
\newcommand{\maxed}[1]{ {#1}_{\text{max}} }	
\newcommand{\atr}{A}

\newcommand{\wbs}{\omega_{\text{BS}}}
\newcommand{\thconf}{\Lambda_{\text{\isconf}}}
\newcommand{\thconfo}{\Lambda_{\text{\isconf}, 0}}
\newcommand{\Thetads}{\Theta_{\text{DS}}}	

\renewcommand{\o}{\varnothing}	
\newenvironment{compactlist}{\begin{minipage}[t]{\linewidth}\begin{list}{$\bullet$}{\leftmargin=0.5em \rightmargin=0em \topsep=0em \parskip=0em \parsep=0em \listparindent=0em \partopsep=0em \itemsep=0pt \itemindent=0em \labelwidth=\leftmargin\labelsep+0.25em}}{\end{list}\end{minipage}} 

\begin{document}

\title{A Plausibility-based Fault Detection Method for High-level Fusion Perception Systems}

\graphicspath{{./Pics/}}

\author{Florian Geissler, Alexander Unnervik, and Michael Paulitsch
\thanks{Paper submitted on June 26, 2020. This work was partially funded by the German Federal Ministry of Transport, Building and Urban Development (BMVI) within the projects KoRA9 (grant No. 16AVF1032A) and Providentia++ (P++) (grant No. 01MM19008). We further thank Soeren Kohnert, Hochschule of Augsburg, for recording highway Radar data.}
\thanks{Florian Geissler and Michael Paulitsch are with Intel Labs, 85579 Neubiberg, Germany. Email: florian.geissler@intel.com}
\thanks{Alexander Unnervik was with Intel Labs, 85579 Neubiberg, Germany, while this research was conducted. He is now with Idiap Research Institute, 1920 Martigny, Switzerland.
}}

\maketitle

\begin{abstract}
Trustworthy environment perception is the fundamental basis for the safe deployment of automated agents such as self-driving vehicles or intelligent robots.
The problem remains that such trust is notoriously difficult to guarantee in the presence of systematic faults, e.g. non-traceable errors caused by machine learning functions.
One way to tackle this issue without making rather specific assumptions about the perception process is plausibility checking. 
Similar to the reasoning of human intuition, the final outcome of a complex black-box procedure is verified against given expectations of an object's behavior. 
In this article, we apply and evaluate collaborative, sensor-generic plausibility checking as a mean to detect empirical perception faults from their statistical fingerprints. 
Our real use case is next-generation automated driving that uses a roadside sensor infrastructure for perception augmentation, represented here by test scenarios at a German highway and a city intersection. The plausibilization analysis is integrated naturally in the object fusion process, and helps to 
diagnose known and possibly yet unknown faults in distributed sensing systems.
\end{abstract}



\section{Introduction}

The verification of a complex environment model without explicit knowledge of -- or access to -- the full processing pipeline is one of the key challenges of modern perception systems.
Causes for incorrect observations can involve electronic system malfunctions of sensors or unintended system behavior, i.e. misperception in the absence of malfunctions such as misuse or incomplete information \cite{Rau2019}.
Notably, object detection further relies more and more on machine learning techniques such as deep neural networks, which are known to be sensitive to particular faults that defy a classical safety assurance (e.g. erroneous training data, ontology gaps, adversarial noise, etc.) \cite{Burton2017}. 
We here refer to a wider class of root causes for the incorrect environment perception of an intelligent sensing device as \textit{perception faults}.
The detection and evaluation of such perception faults is the prerequisite for a safety analysis concerning the user of the environment model, e.g. an automated vehicle (AV), as regulated by standardization bodies such as the ISO 26262 \cite{Iso26262} or ISO/PAS 21448 (SOTIF) \cite{Sotif2019}.

Among quantitative model-based verification approaches \cite{Venkat2003, Avizienis2004}, we see a characteristic trade-off between the complexity of the verification process and the coverage of potential faults: Traditional fault diagnosis techniques, based e.g. on a direct hardware or software redundancy, provide variable sets of high completeness, leading to a potentially high fault coverage. However, a replication of components is neither cost-efficient nor feasible in many scenarios. In contrast to that, analytical redundancies inferred from known functional dependencies on the input variables, or a priori rules, allow to check output variables with greatly reduced complexity. For that reason, \textit{plausibility checks} that verify complex environment variables against a range of expected outcomes are promising candidates for the verification of black-box perception systems \cite{Burton2017}. 
On the other hand, functional relationships are often unknown or ambiguous, resulting in a lack of diagnostic residuals which can diminish the fault coverage. It is therefore of great interest to numerous safety-critical applications if and how such an approach is able to diagnose realistic perception faults.

\begin{figure*}[t]
\centering
\begin{subfigure}{0.7\textwidth}
  \centering
	\includegraphics[width=1.\textwidth]{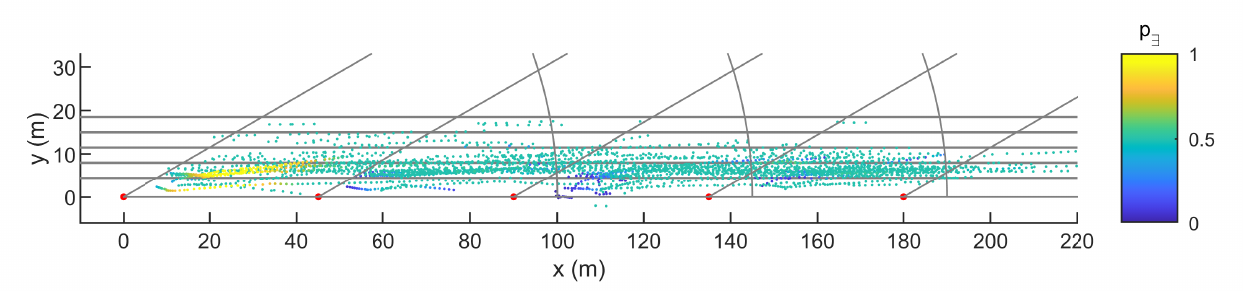}
\end{subfigure}
\begin{subfigure}{0.3\textwidth}
  \centering
	\includegraphics[width=1.\textwidth]{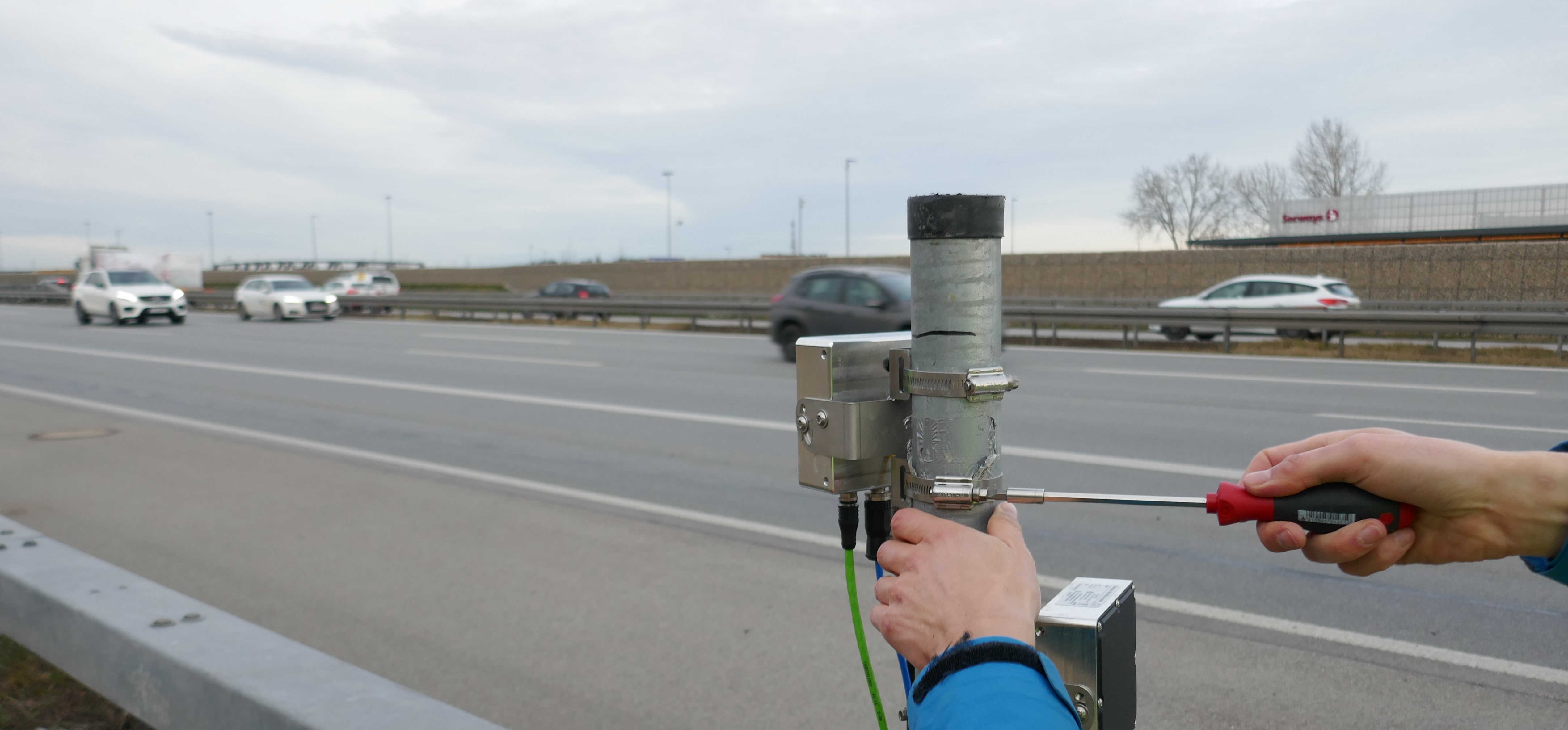}
\end{subfigure}
\caption{An example experiment performed in the KoRA9 project, see also \cite{Geissler2018}. (Left) Object tracks accumulated over time at a highway test segment. Plausibilization can be applied to identify false positive tracks, and reveal inconsistencies in regions of sensor overlaps. (Right) One elementary perception fault is the misorientation of individual sensors at the roadside.}
\label{fig:kora9}

\vspace{0.5cm}
\small
\begin{tabularx}{0.95\textwidth}{p{0.05\textwidth} p{0.18\textwidth} X} 
			\toprule
			Class & Local effect (Error) & Potential fault cause \\ \midrule 
			\textbf{1} & Sensor reports readings with incorrect angles & 
			\begin{compactlist}
			\item \textbf{Misoriented sensor pose} (e.g. poor calibration, air blasts of passing vehicles, accidents, malicious intervention)
			\item Incorrect signal processing
			\end{compactlist} \\ 
			\textbf{2} & Sensor has a higher false positive rate than expected & 
			\begin{compactlist} 
			\item Defect component causing increased thermal noise
			\item \textbf{Incorrect parametrization of algorithmic component} (e.g. tracking or detection threshold)
			\item Incorrect signal processing (e.g. ground plane filter, side lobe artefacts)
			\item (Local) spoofing
			\item Increased signal multipath propagation (e.g. highly reflecting object in sensor field of view)
			\end{compactlist} \\ 
			\textbf{3} & Sensor has a higher false negative rate than expected & 
			\begin{compactlist} 
			\item \textbf{Sensor surface polluted}
			\item Incorrect parametrization of algorithmic component
			\item Incorrect signal processing
			\end{compactlist} \\ \bottomrule
\end{tabularx}

\captionof{table}{Typical safety-critical perception errors and faults, as identified from real smart infrastructure experiments. Faults with similar local effects in the system are combined to fault classes. The selected highlighted faults are injected in simulation in Sec.~\ref{sec:results} to test plausibility-based fault evaluation.}
\label{tab:type_errors}
\end{figure*}
\normalsize

In this article, we investigate the suitability of plausibility checking for the evaluation of perception faults in the context of a distributed sensor network (DSN) for collaborative environment perception. Our use case is a roadside sensor network that supports automated vehicles with complementary high-quality data \cite{Seebacher2019}, a concept also known as \textit{smart infrastructure}, that represents one of the driving forces in the fast-growing market of intelligent transportation systems and smart cities \cite{MarketsandMarkets2018}.
Verification methods as the one studied in this article allow to quantify trust in the integrity of the perception data, and will be required to achieve certifiably safe infrastructure systems in the future (see e.g. \cite{AnnaCarreras2018}).

DSNs typically adopt a \textit{decentralized} tracking architecture, where each sensing device performs a series of processing steps and eventually provides with lists of tracked objects.
The data from multiple sensors are then combined in a central node in the course of high-level sensor data fusion (SDF), or \textit{object fusion}. 
Starting from the architecture of \cite{Aeberhard2011} and the track-to-track (T2T) fusion implementation of \cite{Houenou2012}, we here present a plausibility checking framework that integrates naturally into the object fusion process. Uncertainties are handled within the Dempster-Shafer (DS) theory of evidence \cite{Shafer1976, Challa2004}.
Besides an instantaneous plausibility assessment, we provide with statistical metrics in order to identify systematical faults.
While additional customized plausibility checks can be readily incorporated due to the modular structure of the framework, we here focus on sensor-generic checks that are not limited to a specific modality of the underlying sensor network (most commonly camera, Radar, Lidar), in order to keep the methodology widely applicable.

We test our plausibility-based fault diagnosis in simulation, using the injection of typical perception faults as identified from real-world infrastructure experiments. In particular, we use insights from the German publicly funded projects KoRA9 \cite{BMVI2017} and P++ \cite{BMVI2020}. In KoRA9, environment perception with a Radar-only roadside sensor network is studied at a German highway, see also Fig.~\ref{fig:kora9}. The P++ project on the other hand is set up to monitor a cohesive test area of highway, rural, and urban roads with a variety of sensor modalities including camera and Radar. The Tab.~\ref{tab:type_errors} gives examples of empirical, potentially safety-critical faults that challenge roadside perception systems due to their systematic, erratic, and unsignaled nature. Selected fault instances that are representative of broader classes with similar local effects will be used in this article to identify characteristic signatures.

In summary, the novel contribution of our work is twofold:
\begin{itemize}
\item 
While in \cite{Aeberhard2011} separate modules for T2T and existence fusion are envisioned, we here describe a more efficient combined approach, meaning that plausibility checks and existence estimation are naturally integrated into the object fusion process. We further introduce additional checks that correlate the \textit{track status} of an object with other observed attributes. This allows for a more differentiated and realistic situational assessment.
\item 
We demonstrate that plausibility-based metrics represent a valid strategy to detect realistic perception in the given setup. 
The signatures of three empirical faults are studied in simulation, and characteristic fingerprints are identified. We provide an interpretation of the plausibility metrics with respect to the underlying source of implausibility.
\end{itemize}

The article is structured as follows: In Sec.~\ref{sec:sota} we review related literature, before we present our model in Sec.~\ref{sec:model}, including the process integration and the various plausibility checks. Next, Sec.~\ref{sec:faults} describes the method for the detection of perception faults. The respective plausibility signatures are analyzed in Sec.~\ref{sec:results} for two different scenarios, before we conclude in Sec.~\ref{sec:summary}.

\section{Related work}
\label{sec:sota}

First, it is key to note the difference between measurement uncertainty, plausibility, and risk. The former refers to the statistical inaccuracy of a sensor observation, and its propagation throughout the system. Plausibility represents a notion of trust in a hypothesis (e.g. "object exists"), that is established by checking a piece of information against expectations based by a priori knowledge or consistency with other sources. As given in Sec.~\ref{sec:model}, the plausibility of an existence hypothesis can be conveniently mapped onto a \textit{probability of existence} measure. Risk on the other hand refers to an event, which can occur with a certain likelihood and a given impact on a user. Probabilities of existence and uncertainties are valuable inputs to a risk analysis, but can not provide with any context or user impact. In this article, we are not concerned with the explicit evaluation of measurement uncertainty or risk, but with plausibilization.

Different hierarchies of plausibility checking are to be distinguished, the authors of \cite{Versmold2006} define for example three categories of increasing complexity: i) Single signal monitoring, ii) redundancy-based checking, and iii) model-based checking. Interesting examples for the latter category are e.g. the plausibilization of trajectory curvature against possible lateral acceleration \cite{Yavvari2018}, or unphysical overlaps of object shapes \cite{Bissmeyer2010}.
Among the established mathematical methods to quantify confidence, in particular Bayesian belief networks and the DS theory have proven to be appropriate tools \cite{Graydon2017}.

In the domain of safety for automated driving, plausibility checking methods have been applied to verify the environment perception of (overlapping) in-vehicle sensors -- including camera, Lidar, and Radar modalities -- and to infer a probability of existence metric \cite{Maehlisch2007, Aeberhard2011, Aeberhard2017}. Most of the applied checks there are single-signal and redundancy checks. 
This assessment can take place at different stages of the SDF, and the checks take different forms depending on the available features.
High-level object fusion architectures with DS plausibilization have for example been studied in \cite{Maehlisch2007, Aeberhard2011}. Here, the position of an object is for example checked against the position of other objects to identify implausible occlusions, and a non-uniform belief is assigned across a sensor's field of view (FoV). 
Another body of literature harnesses plausibility checking to address secure vehicle-to-everything (V2X) communication, see e.g. \cite{Bissmeyer2010, Obst2015, Ambrosin2019}. Information that is transmitted to a vehicle from external sources, such as the infrastructure or other vehicles, is plausibilized to detect misbehavior or spoofing. A basis for verification are the parameters contained e.g. in the format of a basic safety message (BSM), including position, speed, acceleration or object dimensions \cite{Yavvari2018}, but also the sensing and communication range.

\section{Model}
\label{sec:model}

\subsection{Object representation}

\begin{figure*}[tbp]
\centering
\includegraphics[width=1.\textwidth]{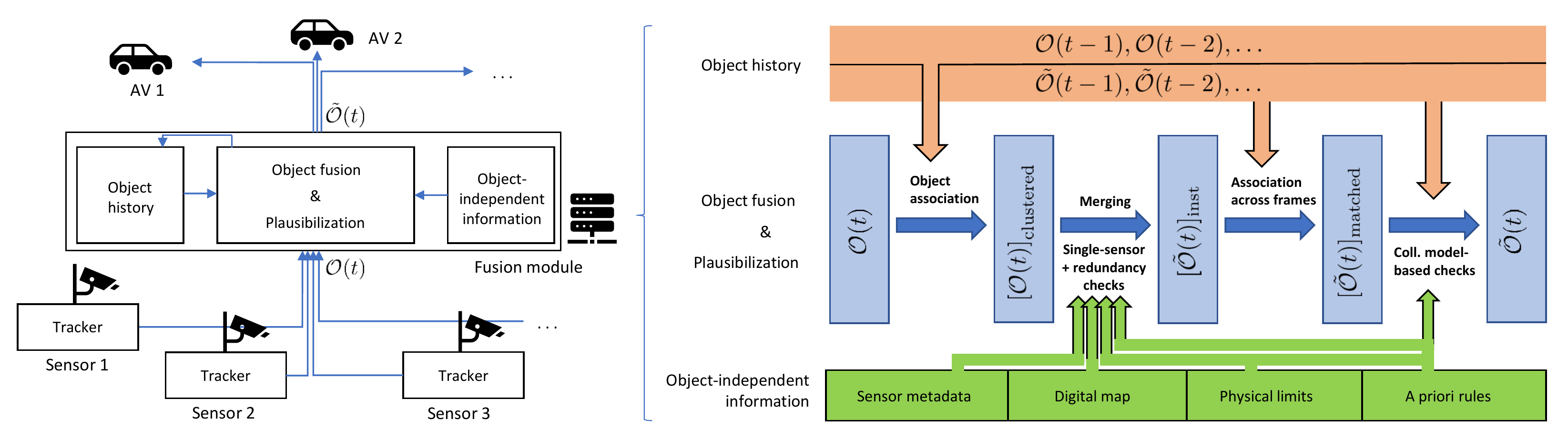}%
\caption{(Left) System sketch of the decentralized tracking architecture for collaborative perception. (Right) Schematic of the fusion module, performing combined object fusion and plausibilization.}
\label{fig:sysAll}
\end{figure*}

The Fig.~\ref{fig:sysAll} gives a schematic overview of a high-level fusion system. Objects tracked by the individual sensors are passed to a central \textit{fusion module}. In the highlighted infrastructure use case, this component runs for example on an edge cloud server at the roadside. 
We here focus on an integrated model of object fusion and plausibilization as the core functionality of this module. 

Let us define a tracked object reported by an individual sensor as a structured list $\Ostate$, while $\Olistloc$ is the \textit{local object list}, i.e. the set of all individually registered objects at a common time step,
\begin{align}
\Ostate & =\left(i, n, t, \xstate, \bm{P}, \ddim, \trstatus, \ldots \right) \equiv \Ostate^i_n(t),
\label{eq:localObj} \\
\Olistloc(t) & = \{ \{ \Ostate^i_n(t)| n=1,2,\ldots \} | i=1,2,\ldots  \}.
\label{eq:defOlistloc}
\end{align}
The minimal required object attributes in our model are: $i$ sensor index, $n$ track identifier (ID), $t$ time stamp, $\xstate$ object state, $\bm{P}$ state covariance matrix, $\ddim$ object dimensions, and $\trstatus$ tracking status information. Any object can be uniquely identified by $(i, n, t)$. For notational ease, we only give selected attributes if clear from the context (e.g. $\Ostate^i$ refers to a given object reported by a sensor $i$, $\xstate_{\Ostate^i} \equiv \xstate^i$, etc.). 
%
We specify
\begin{equation}
\label{eq:defx}
\begin{split}
\xstate &= (x, y, z, v_x, v_y, v_z)', \\
\ddim &= (L,W,H, h), \\
\trstatus & = \left(\sco, \isconf, \iscoast, \ldots \right).
\end{split}
\end{equation}

Object position and velocity are represented by $x,y,z$ and $v_x, v_y, v_z$, respectively. The dimension entails length ($L$), width ($W$), height ($H$), and heading ($h$) of the perceived object. 
In our simulation, we adopt a sequential multi-object tracking (MOT) scheme  -- different implementations are possible and do not affect the general model -- where the track score $\sco$ is a log-likelihood ratio \cite{Blackman1999} that increases (decreases) with each incoming (missing) update. 
The binary value $\isconf$ indicates whether the object track is a confirmed ($\isconf=1$), or a tentative ($\isconf=0$) track. 
Furthermore, the binary $\iscoast$ incorporates information about the coasting status, i.e. whether the object track is updated at the current time step ($\iscoast=0$) or not ($\iscoast=1$). 

Next, a \textit{system object} $\Obstate$ is defined to be a global tracked object, that can be observed by multiple sensors at the same time, and therefore can have multiple varying \textit{local representations}. 
Similar to Eq.~\ref{eq:defOlistloc} we define the \textit{system object list} $\Olistsys$ at a common time $t$,
\begin{align}
\Obstate &=\left(n', t, \xstate, \bm{P}, \ddim, \trstatus, \pp, \spp, \ldots \right) \equiv \Obstate_{n'}(t),
\label{eq:sysObj} \\
\Olistsys(t) &= \{ \Obstate_{n'}(t)| n'=1,2,\ldots \}.
\label{eq:defOlistsys}
\end{align}
Here, $\pp$, $\spp$ are the respective probability existence and existence uncertainty metrics derived in Sec.~\ref{sec:DST}, using the concepts of belief and plausibility. 
The goal of the object fusion with integrated plausibility checking is the transformation $\Olistloc(t) \xrightarrow{} \Olistsys(t)$, under consideration of temporal continuity.
In the process, the metrics $\pp$ and $\spp$ are assigned to give a quantifiable estimate for trust in the observation.

\subsection{Fusion module}
\label{sec:T2Tfusion}

The integral part of this process is track-to-track fusion \cite{Blackman1999}, which combines the dynamic state variables of multiple tracks. Furthermore, object fusion merges features like dimensions, headings etc. 
We encounter the following steps (see Fig.~\ref{fig:sysAll}): 
\begin{itemize}
\item{\textbf{Object association}}
Synchronous and confirmed local representations in $\Olistloc$ are grouped to \textit{track clusters}, $\left[\Olistloc(t)\right]_{\text{clustered}}$. We adopt the T2T distance calculation and the global nearest neighbor (GNN) association of \cite{Houenou2012}, which can include the history of local representations for better association robustness. 

\item{\textbf{Merging}}
All local object representations within one cluster are merged to a single system object \cite{Houenou2012}. The result of this fusion process is a system object list $\left[\Olistsys(t)\right]_{\text{inst}}$ (for \textit{instantaneous}), with arbitrary IDs.
\item{\textbf{Association across frames}}
We associate the unique object labels from the previous and the current time frames via another GNN association, so that persistent objects keep the same unique ID in the matched list $\left[\Olistsys(t)\right]_{\text{matched}}$. Dimensions and existence estimations can be averaged across recent frames for improved robustness.
\end{itemize}

Our model now integrates various plausibility checks in the object fusion process to evaluate the quality of the sensor data.
Extending the paradigm of \cite{Versmold2006}, we have three different types of checks, see Fig.~\ref{fig:sysAll} and Sec.~\ref{sec:BBA} - \ref{sec:modelchecks}:
\begin{itemize}
\item{\textbf{Single-sensor checks}}
are based on the attributes of local representations from a \textit{single} sensor and result in the basic belief assignment of the DS framework.
An attribute is here checked against predefined thresholds (signal-based) or known relations to other attributes (model-based). 
\item{\textbf{Redundancy checks}}
compare local object representations from \textit{different} sensors that refer to the same system object, as identified by object association. Such redundancy checks are applied after the clustering, and before the merging step. Here, they are represented by the DS combination, together with the registration of misses and unexpected observations.
\item{\textbf{Collaborative model-based checks}}
aim to verify the final outcome of the fusion process against predefined rules about the expected behavior of system objects. 
This additional processing step transforms the output of the frame-association process $\left[\Olistsys(t)\right]_{\text{matched}}$ into the desired output $\Olistsys(t)$.
Such checks may be based on instantaneous interrelations among merged object attributes, or on the evolution of merged attributes over time. In this work, we present two possible checks that can be used to tackle false positives or spoofed detections (see Sec.~\ref{sec:modelchecks}). Various further checks of increasing complexity can be envisioned at this point, including e.g.  traffic-rule-compliant trajectory checking, or self-learned motion patterns.
\end{itemize}

The process diagram of Fig.~\ref{fig:sysAll} further illustrates which information beyond the sensor object data is required to perform the plausibility checks described in our model. 
Single-sensor (and partially the redundancy) checks build upon the knowledge of sensor metadata such as the sensor field of view and mounting locations, or the expected road boundaries of a digital map. Physical limits of variables and predefined rules help to establish additional plausibility for all three types of checks.

\subsection{Single-sensor checks}
\label{sec:BBA}

Compared to a Bayesian-based object existence estimation, the well-studied Dempster Shafer theory of evidence \cite{Shafer1976, Challa2004, Dezert2012} postulates an extended probability space of $\Thetads=\{ \o, \exists, \nexists, \unk \}$, where $\o$ is the null set, $\exists$ and $\nexists$ are the propositions that a given object exists or does not exist, respectively, and the element $\unk$ represents the possibility that the existence of the object is unknown.
Basic belief assignments (BBA) give rise to belief masses that indicate the degree of belief in a proposition according to a given sensor. 
Explicitly, $m_i(X \Obstate)$ denotes the belief mass for the proposition $X \in \Thetads$ of $\Obstate$ based on the local observation $\Ostate^i$ of sensor $i$.

The following sensor-generic plausibility checks define our BBA,
\begin{equation}
\label{eq:BBA}
\begin{split}
 m_i({\exists \Obstate}) &= \ptrust{i}\ \pfov{}{\Ostate^i}\ \pocc{}{\Ostate^i}\\
		& \times \pex{}{\Ostate^i}\ \pdm{}{\Ostate^i}\ \pval{}{\Ostate^i},  \\
 m_i({\nexists \Obstate}) &= \ptrust{i}\ \pfov{}{\Ostate^i}\ \pocc{}{\Ostate^i}\\
		& \times \left[1-\pex{}{\Ostate^i}\ \pdm{}{\Ostate^i}\ \pval{}{\Ostate^i}\right],  \\
 m_i({\unk \Obstate}) &=1- \left[m_i({\exists})+ m_i({\nexists}) \right].
\end{split}
\end{equation}

Note that despite the formal equivalence to some of the models of \cite{Maehlisch2007, Munz2009, Aeberhard2011a, Aeberhard2017}, the BBA factors in Eq.~\ref{eq:BBA} take a different form due to the fact that we study tracked objects instead of detections, which leads to  important additional interrelations of the attributes of $\Ostate^i$. 
The plausibility factors $p$ can be grouped into two categories: A reduced value of $\pany{trust},\pany{FoV}, \pany{occ}$ will support ignorance of the system object, while a low weight on $\pany{ex}, \pany{dm}, \pany{val}$ favors the proposition of non-existence. The individual factors defining the BBA are explained in more detail in the following.

First, $\ptrust{i}$ is an object-independent trust factor for the sensor $i$, which can be interpreted as its typical object detection capability \cite{Aeberhard2017}.

The factor $\pany{FoV}$ checks whether the reported object is in the three-dimensional FoV, here defined as $\fv^i=(r^i, \omega^i, \psi^i)$, where $r$ is the range, $\omega$ and $\psi$ the horizontal and vertical view angle, respectively.  
Given that at least one point across the body of an extended object (here we check the eight corners of the bounding box, object center, and two additional points in the middle front and rear of the bounding box) is found to be within the expected FoV, we denote $(\xstate^i, \ddim^i) \in \fv$. Otherwise, a distance metric $\text{D}$ is applied to measure the closest distance from the object center $\xstate^i$ to the FoV. For example, $\text{D}(\xstate^i, r^i)$ is the difference of the radial distance of $\xstate^i$ and the sensor range, and analogous for $\omega, \psi$.
We then define
\begin{equation}
\pfov{}{\Ostate^i}=
\begin{cases}
1 &\text{if $(\xstate^i, \ddim^i) \in \fv^i$},\\
e^{-\sum_{n=1}^3 \text{D}(\xstate^i, \fv^i_n)/(\fv^i_n/2) }&\text{otherwise}.
\end{cases}
\label{eq:defPfov}
\end{equation}
This choice of weighting is motivated by two additional considerations: i) Tracked objects can leave the FoV while coasting, in which case it is legitimate to have $\pany{FoV}>0$;  ii) Tracks far away from the original FoV should be suppressed, whether coasting or not (otherwise unrealistic, e.g. spoofed tracks that pop up at an arbitrary remote location and start coasting would be supported).

Next, $\pany{occ}$ verifies if an object is rightfully observable in relation to other objects perceived by the same sensor. 
If at least one of the checked surface points of $\Ostate^i$ is in the line of sight of sensor $i$, given all other confirmed observations of $i$, we denote $(\xstate^i, \ddim^i) \in \Los^i(\{\Ostate^i\})$. 
If that this is \textit{not} the case, and at the same time the track is not coasting, the object is occluded and the belief in this observation is forfeited. 
A track that is coasting in the shadow of another object will not be affected. 
\begin{equation}
\pocc{}{\Ostate^i}=
\begin{cases}
0 &\text{if $(\xstate^i, \ddim^i) \notin \Los^i(\{\Ostate^i\})$ $\wedge$ $\iscoast^i =0$},\\
1 &\text{otherwise}.
\end{cases}
\label{eq:defPocc}
\end{equation}

Furthermore, $\pex{}{\Ostate^i}$ is the existence probability of the object as estimated by sensor $i$. Here, we infer $\pany{ex}$ directly from the track score \cite{Blackman1999, Musicki2006},  
\begin{equation}
\pex{}{\Ostate^i}=1/\left( 1+e^{-\alpha\ \text{score}^i + \beta} \right)
\label{eq:defPex},
\end{equation}
where $\alpha$ and $\beta$ are tunable coefficients to control the slope of the sigmoid function. We here chose $\alpha$ and $\beta$ such that a newly initialized, tentative track starts with $\pany{ex}= 0.9$, and increases to $\pany{ex}= 0.99$ if the score reaches the confirmation threshold.
A detection missed by sensor $i$ will be handled as $\pex{}{\Ostate^i}=0$, leading with Eq.~\ref{eq:BBA} to $m_i(\exists\Obstate)=0$, $m_i(\nexists\Obstate)=\ptrust{i}$, and $m_i(\unk\Obstate)=1-\ptrust{i}$.

The factor $\pdm{}{\Ostate^i}$ compares the position of an object to the expected road boundaries. Let $dm$ be a digital grid map with binary values corresponding to road $(1)$, and non-road ($0$) squares. For each object, the metric $\text{D}(\xstate^i, dm)$ then calculates the Euclidean distance from the measured position $\xstate^i$ to the closest road square. As a normalization factor we use a typical lane width \cite{Verlag2008} of $W_{\text{lane}}=3.5m$,
\begin{equation}
\pdm{}{\Ostate^i} = e^{- \text{D}(\xstate^i, dm)/W_{\text{lane}} }.
\end{equation}

Finally, we have the signal value factor $\pany{val}$ which compares object attributes against physically possible limits. In particular, we check the vertical position relative to the road, dimensions, and absolute velocity, 
\begin{equation}
\pval{}{\Ostate^i} =  \exp\left(- \sum_{\atr \in \{z, W, L, H, |v|\}} [\atr^i,\maxed{\atr}]_+/\maxed{\atr} \right).
\end{equation}
Here we denote $[A, B]_+ = \theta(A - B) (A - B)$ as the positive real difference, where $\theta(\ldots)$ is the Heaviside function. The predefined maximum limit of an attribute $\atr$ is represented by $\maxed{\atr}$, and we choose an empirically parameter set of $\maxed{z}=3m$, $\maxed{W}=5m$, $\maxed{L}=25m$, $\maxed{H}=5m$, and $\maxed{|v|}=80m/s$.

\subsection{Redundancy checks}
\label{sec:DST}

To obtain an overall notion of plausibility of $\Obstate$, the belief masses of the individual sensors referring to the same system object are fused  -- note $\oplus$ -- with the help of the DS combination rule \cite{Dezert2012}. Assuming a total number of $N$ sensors, we arrive at the fused belief masses $m_F$,
\begin{equation}
m_F(X \Obstate) \equiv [m_1 \oplus m_2 \oplus \ldots \oplus m_N](X \Obstate).
\label{eq:DScombo3}
\end{equation}

Iterating through all sources, any sensor that makes a local observation of $\Obstate$ can possibly generate non-zero belief masses for all
propositions $X \in \Thetads \backslash \o$. We register \textit{unexpected} observations when a reported object is outside the non-occluded FoV of the reporting sensor, and at the same time is non-coasting. These situations will produce a high belief in the proposition $ m_i(\unk \Obstate)\simeq 1$.
If a given sensor $i'$ has \textit{no} local representation in the respective cluster, it matters whether it failed or was unable to perceive the object. Therefore, we check if $\Obstate$ (as given by the available reports of other sensors) is in the non-occluded field of view of sensor $i'$, in which case we register a \textit{miss} for $i'$, along with the mass $m_{i'}(\exists \Obstate)= 0$ (see Sec. ~\ref{sec:BBA}).
Further, a miss is as well registered (without change of the masses) for a sensor that fails to update an object in its non-occluded FoV, as can be inferred from a positive coasting status.
Finally, if $i'$ has no representation but $\Obstate$ is outside its FoV or occluded, $i'$ is an irrelevant contributor, and we have $m_{i'}(\unk \Obstate)=1$.
Such sensors do not have any impact under the DS combination rule. Unexpected observations and misses will be used for a statistical fault analysis (see Sec.~\ref{sec:faults}).


\subsection{Collaborative model-based checks}
\label{sec:modelchecks}

We describe two additional, sensor-generic checks that help to verify the integrity of high-level perception data based on collaborative knowledge.
More complex or sensor-specific model-based checks -- e.g. related to object classes or trajectories -- can be incorporated the same way, and are envisioned for further research.

\textbf{Plausible observation history.} 
We check whether an \textit{increase} of belief in the object existence across two subsequent time steps is justified. This is not the case if there is \textit{no} current observation update obtained from \textit{any} sensor (i.e. if a system object has global coasting status $\iscoast=1$). 
The existence probability and uncertainty of a coasting object then have to be corrected appropriately, with the belief increment $\Delta_{\hist}$ being transferred to the ignorance hypothesis,
\begin{equation}
\label{eq:Delta_hist}
\begin{split}
& \Delta_{\hist, \exists}(\Obstate(t)) = -\left[ m_F(\exists \Obstate(t)), m_F(\exists \Obstate(t-1))\right]_{+}\times \iscoast(t),  \\
& \Delta_{\hist, \nexists}(\Obstate(t)) =  0,\\
& \Delta_{\hist, \unk}(\Obstate(t)) = - \Delta_{\hist, \exists}(\Obstate(t)) .
\end{split}
\end{equation}
This check will prevent objects that have already left the field of view, or spoofed objects with an alleged coasting status to become more relevant.

\textbf{Dimension-velocity check.}
The type of an object is related to its physical speed limitations. Let us here distinguish the two classes of vulnerable road users (VRU, i.e. pedestrians, bicycles, etc.), and other objects (vehicles, trucks, etc.). Since not all sensor modalities can reliably identify object classes, we use the object dimensions for a basic type estimate: VRUs are characterized by the combination of small width and length extensions, and low speed, as parametrized here by the empirical upper thresholds $\lvru=2m$ and $\vvru=20m/s$. 
Small and fast objects will be considered implausible as a consequence, and full ignorance is assumed ($\theta$ is the multi-dimensional Heaviside function),
\begin{equation}
\label{eq:Delta_dimvel}
\begin{split}
&\Delta_{\dimvel, \exists}(\Obstate) = - m_F(\exists \Obstate)\ \theta(\lvru - W, \lvru - L, |v| - \vvru), \\
&\Delta_{\dimvel, \nexists}(\Obstate)  = 0,\\
&\Delta_{\dimvel, \unk}(\Obstate)  = - \Delta_{\dimvel, \exists}(\Obstate).
\end{split}
\end{equation}
Note that object occlusions can temporarily reduce the effectively observed object dimensions, and lead to a low-plausibility assessment in Eq.~\ref{eq:Delta_dimvel}. However, we expect this effect to be short-lived and statistically uniform across all sensors of the network, such that no individual fault patterns will emerge due to this check.

\textbf{Final combined metric.} 
The corrections from all model-based checks are combined as

\begin{equation}
\label{eq:Delta_comb}
\bm{m}_F(\Obstate) \to \left[\bm{m}_F(\Obstate) + \bm{\Delta}_{\hist}(\Obstate) + \bm{\Delta}_{\dimvel}(\Obstate)\right]_0^1,\\
\end{equation}
with $\bm{m}_F(\Obstate) = (m_F(\exists \Obstate), m_F(\nexists \Obstate), m_F(\unk \Obstate))$ and $\bm{\Delta}(\Obstate) = (\Delta_{\exists}(\Obstate), \Delta_{\nexists}(\Obstate), \Delta_{\unk}(\Obstate))$. In case the accumulated increments exceed the correct belief mass interval bounds, we reestablish them by the mapping $[\ldots]_0^1$ applying subsequent shift and renormalization operations. 
The DS framework defines the quantities of overall belief and plausibility \cite{Dezert2012}, however, we here work instead with the related object existence probability $\pp$, and belief interval $\spp$, which can be retrieved from the fused belief masses by the pignistic transformation \cite{Aeberhard2011a}
\begin{equation}
\left(\begin{matrix}
    \pp(\Obstate)       \\
    \spp (\Obstate)       
\end{matrix}\right)
=
\left(\begin{matrix}
    1       & \frac{1}{2}\\
    0       & \frac{1}{2}
\end{matrix}\right)
\left(\begin{matrix}
    m_F(\exists\Obstate)      \\
    m_F(\unk \Obstate)      
\end{matrix}\right).
\label{eq:ppspp}
\end{equation}

With $\pp$, a Bayesian-like probability of the existence of $\Obstate$ from all observed evidence is recovered. On the other hand, $\spp$ can be interpreted as an epistemic uncertainty measure, since it is directly proportional to the weight of the ignorance proposition $m_F(\unk \Obstate)$.

\section{Fault analysis} 
\label{sec:faults}

\subsection{Addressed faults}

For the roadside sensor network at hand, a service failure occurs if the system does not deliver data of sufficient information quality to safely support AVs on the road.
This can happen in the form of various failure modes, commonly categorizing failures from the view points of domain, detectability, consistency, and consequences on the environment \cite{Avizienis2004}. We here focus on system failures that can not be detected easily with component monitoring, but leave statistical signatures to be studied in the plausibility metrics.
Therefore, our addressed failure mode is erratic (data delivery does not halt but is silently corrupted, or mistimed), unsignaled, systematic, and catastrophic (data corruption can be safety-critical).
Importantly, in this article, we adopt a single fault hypothesis. This is justified by the assumption that our analysis, possibly in combination with additional sensor diagnostics, allows for a quick fault detection, that renders the occurrence of multiple faults in that very time interval unlikely. 
Perception faults that were found to be relevant in roadside sensing experiments and testbeds, and that all manifest in the above failure mode, were listed in Tab.~\ref{tab:type_errors}. 
%
The highlighted items there are chosen for further simulation studies in the following section, to investigate their respective fingerprints in the plausibility metrics. 
Our analysis will however be representative of a wider spectrum of faults that belong to the same class, and thus manifest themselves in a similar form.

\begin{figure*}[tbp]
\center
\includegraphics[width=\textwidth]{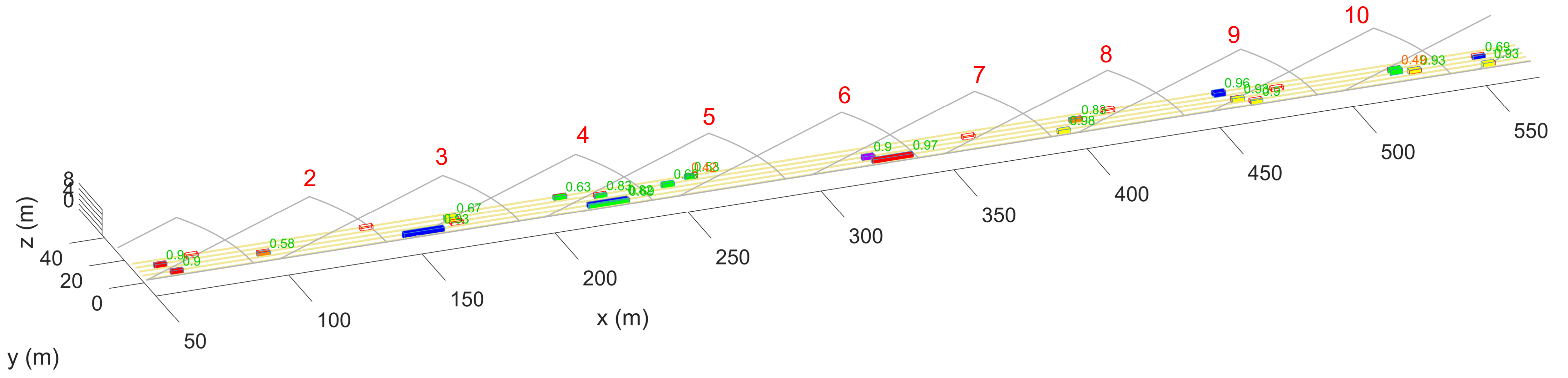} 
\caption{Scenario of a four-lane highway segment monitored by a network of Radar sensors. An existence estimate is assigned to each observed object. Sensor indices are shown in red color.}
\label{fig:setup_simul1}
\end{figure*}

\subsection{Statistical evaluation}

Systematic faults as the ones given in Tab.~\ref{tab:type_errors} alter sensor readings over the course of many sensing cycles, such that we expect statistically relevant deviations from a baseline scenario. The goal of the fault evaluation process is to identify: i) which of the sensors is faulty (assuming single failures), and ii) what fault or fault class is present.
To that end, we inject the selected faults in separate setups, and consider three characteristics for a statistical interpretation: The miss ratio (MR) per sensor, the unexpected observation rate (UOR) per sensor, and most importantly the  probability of existence of an object (involving all sensors, see Eq.~\ref{eq:ppspp}), here in relation to the observed position in the monitored region.
For the former two, we define for a given time and a given sensor,
\begin{equation}
\label{eq:MRUOR}
\begin{split}
 \tMR&=\frac{\text{misses}}{\text{misses + all observations}}, \\
\tUOR&=\frac{\text{unexpected observations}}{\text{all observations}}.
\end{split}
\end{equation}

The above metrics are evaluated at each time step, however, it is important to note that the corresponding variables are not independent due to the nature of the underlying object tracking. For example, an object with a low track score causing a low plausibility at a given time step will likely have a low plausibility in the next time step as well, since the track score only changes gradually.
To make the quantities quasi-iid variables, either of the two following steps can be applied: i) The setup is run various times with independent random distributions of the objects, where the respective signatures are averaged out across one individual scenario. ii) One long scenario run is split into quasi-independent time intervals, where each interval is long enough such that the scene changes significantly from one interval to another. The plausibility metrics within one time interval are averaged out, and the statistics is performed over the resulting iid interval means, assuming a normal distribution. The data basis for our fault analysis is therefore the averages and errors of the various interval means.
Using standard methods \cite{Trivedi2016}, every data point is presented with its $95\%$ confidence interval (CI) in the figures of Sec.~\ref{sec:results}, and a fault is diagnosed if the CI of a interval mean does not overlap with the CI of the baseline scenario (no injected fault).
For MR and UOR, a statistical baseline without faults is established by performing an additional cross-sensor weighted-least-squares average.

In the following section, we analyze three selected fault mechanisms in two different smart infrastructure scenarios motivated by real experiments. The first setup is a highway segment monitored by a network of overlapping Radar sensors, where various vehicle types move on very structured trajectories on multiple lanes. Objects moving in parallel on different lanes occlude each other frequently. 
In the second setting, we explore an urban intersection monitored by two Lidar sensors with high accuracy. This setup includes not only various vehicle types but also cyclists and pedestrians, that all accelerate, stop, and take turns. 
Each temporal interval in our simulations has a sample rate of $0.1s$ and a duration of $5s$, which we found to be a sufficient duration for the highly dynamic scene to change significantly.


\section{Results}
\label{sec:results}
\subsection{Scenario 1: Radar highway infrastructure}

\begin{figure*}[tbp]
    \centering
		\begin{subfigure}{0.33\textwidth}
			\centering
			\includegraphics[width=1\textwidth]{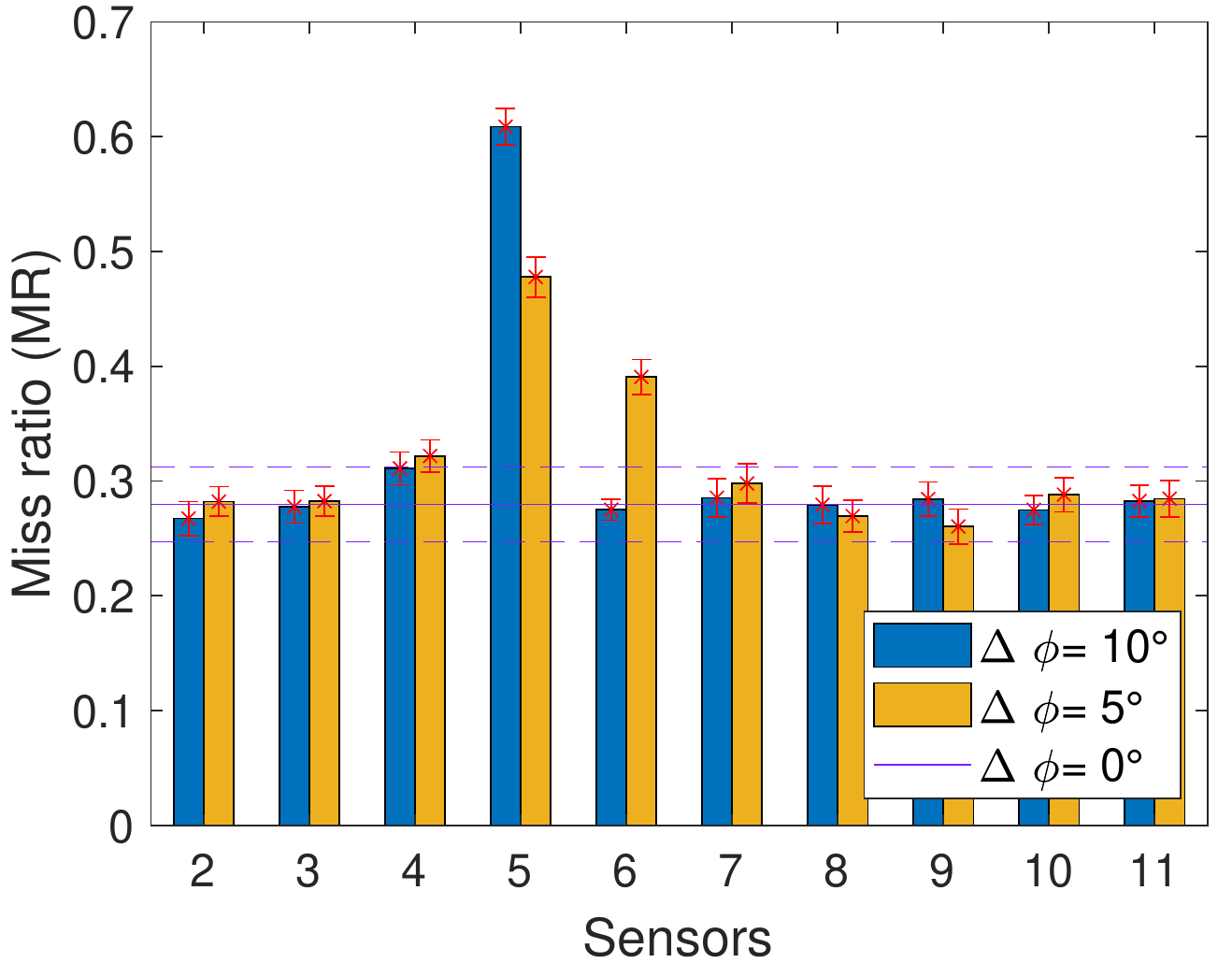}
		\end{subfigure}%
		\begin{subfigure}{0.33\textwidth}
			\centering
			\includegraphics[width=1\textwidth]{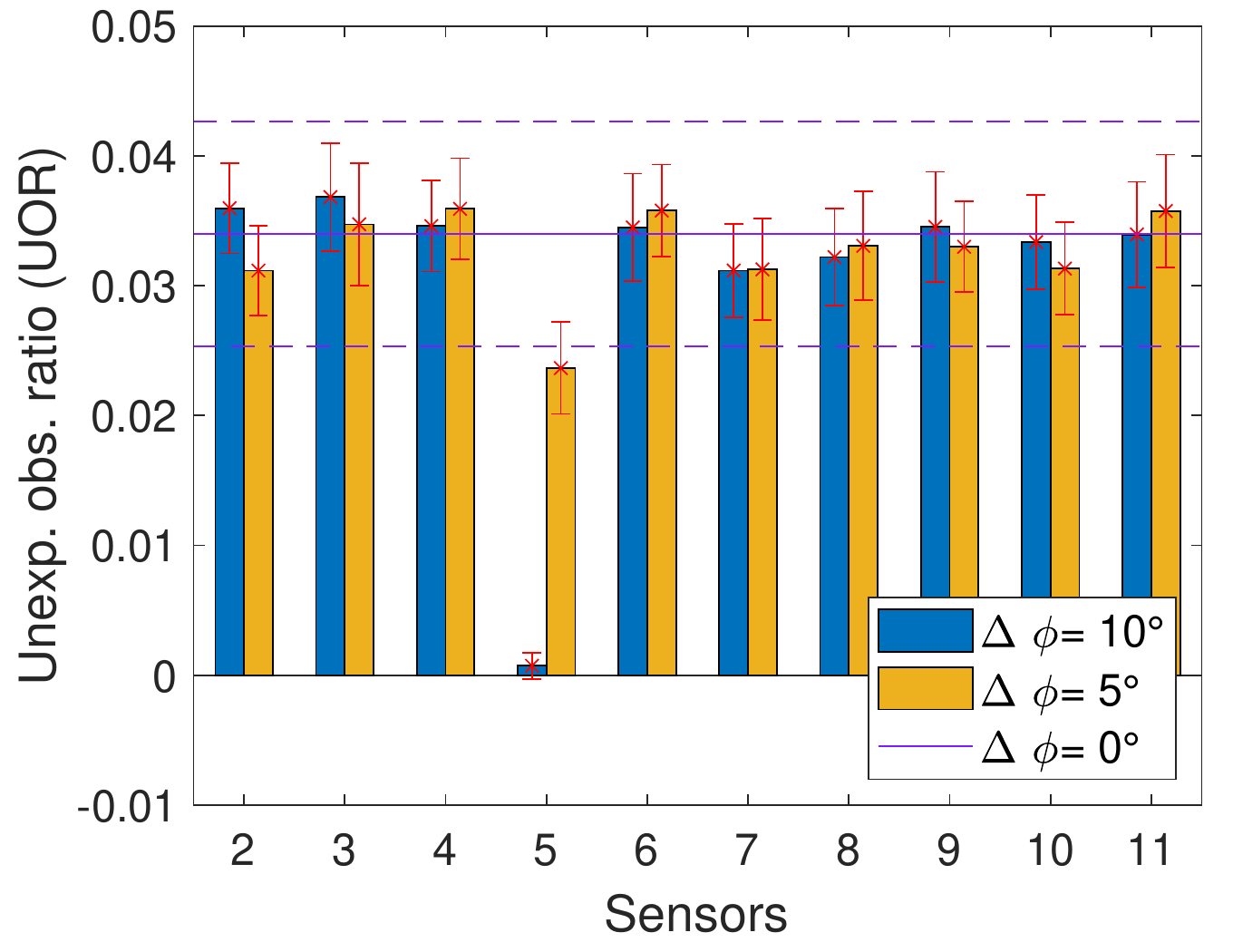}
		\end{subfigure}%
		\begin{subfigure}{0.33\textwidth}
			\centering
			\includegraphics[width=1\textwidth]{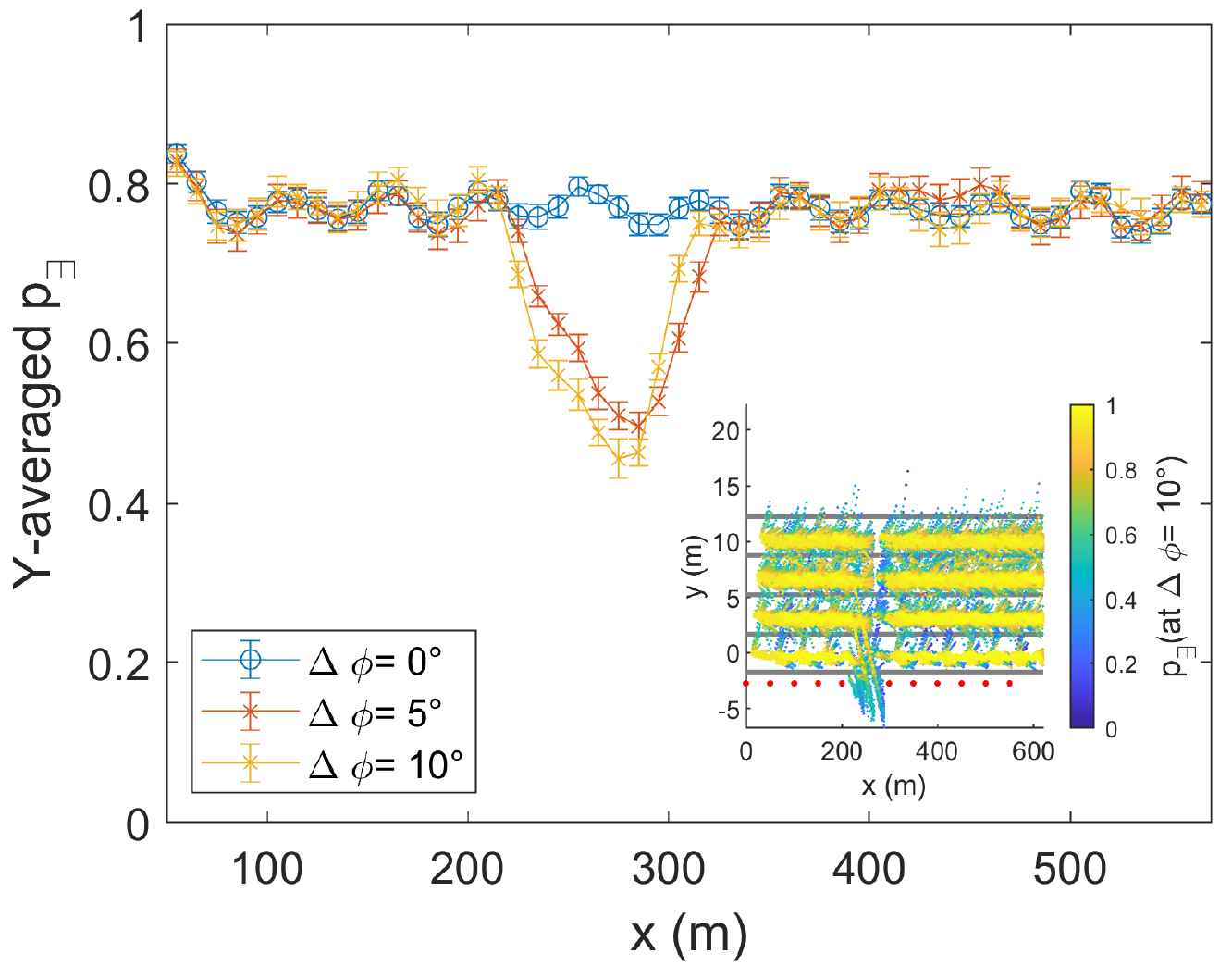} 
		\end{subfigure}%
\caption{Plausibility signatures of the highway scenario of Fig.~\ref{fig:setup_simul1} with injected azimuthal misorientation fault (at sensor $5$). Due to the rotated track headings, already small variations $\Delta \phi$ of a few degrees lead to a characteristic fingerprint in all considered metrics MR, UOR, and $\pp$.}
\label{fig:misError}
\end{figure*}

\begin{figure*}[htbp]
    \centering
		\begin{subfigure}{0.33\textwidth}
			\centering
			\includegraphics[width=1\textwidth]{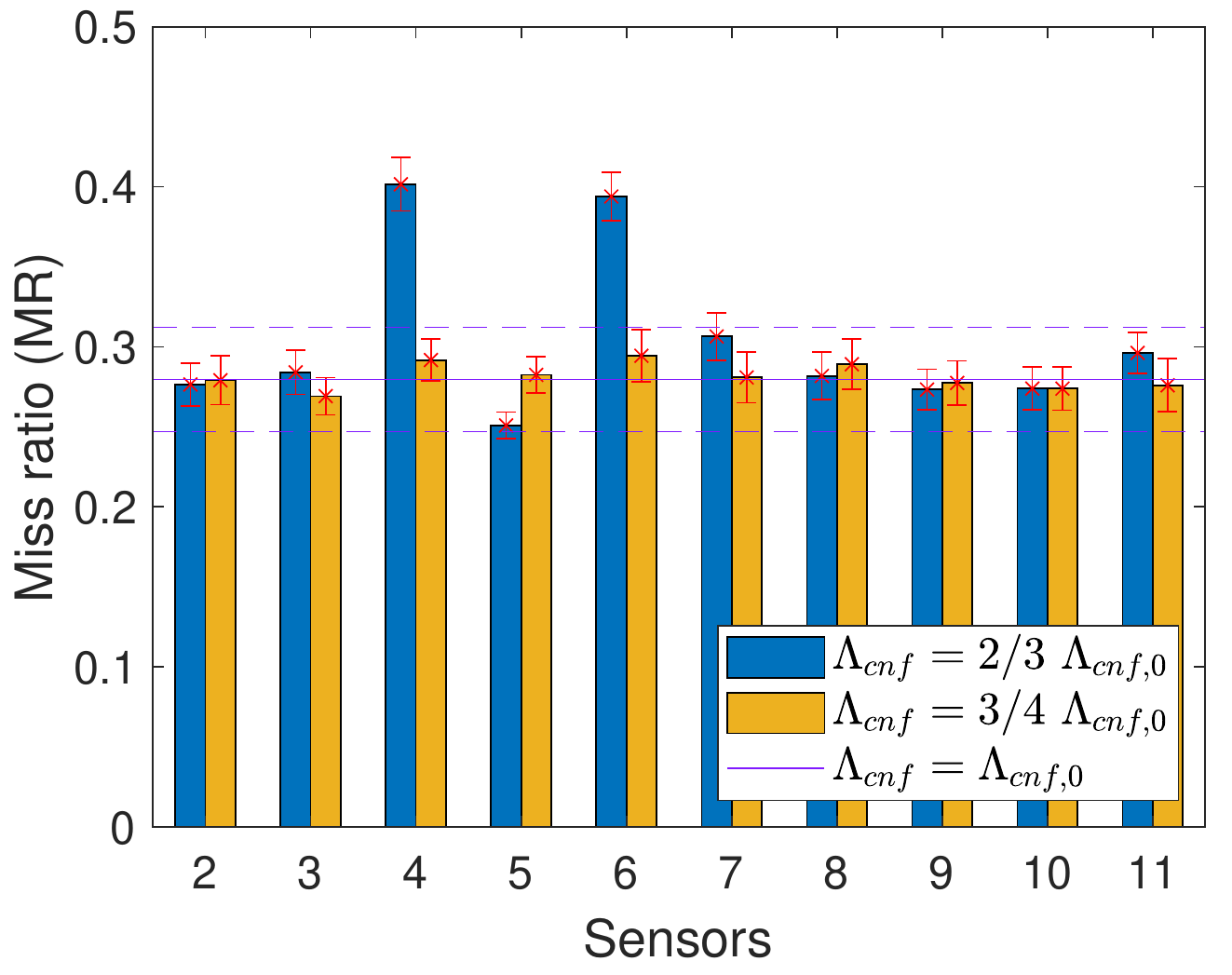}
		\end{subfigure}%
		\begin{subfigure}{0.33\textwidth}
			\centering
			\includegraphics[width=1\textwidth]{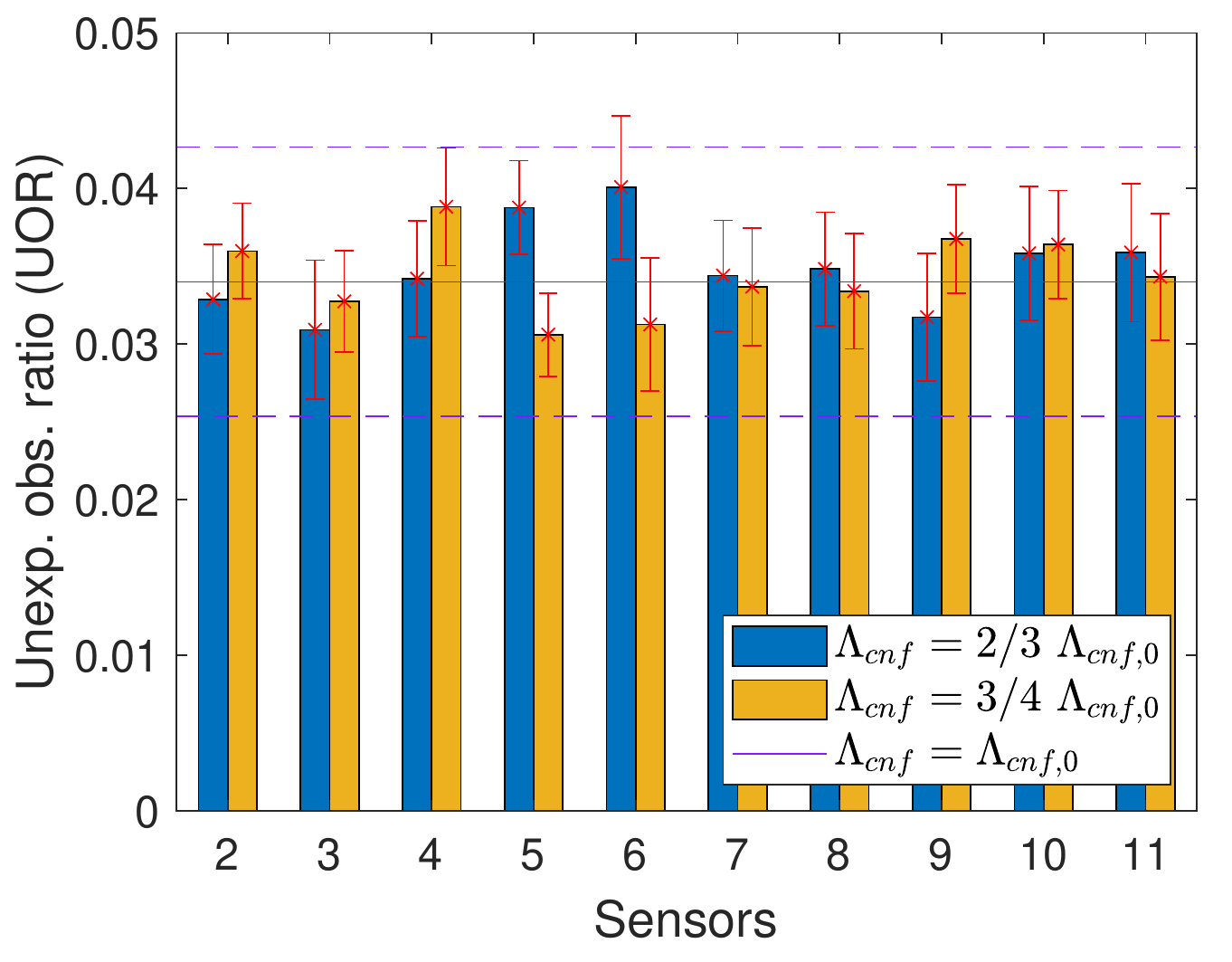}
		\end{subfigure}%
		\begin{subfigure}{0.33\textwidth}
			\centering
			\includegraphics[width=1\textwidth]{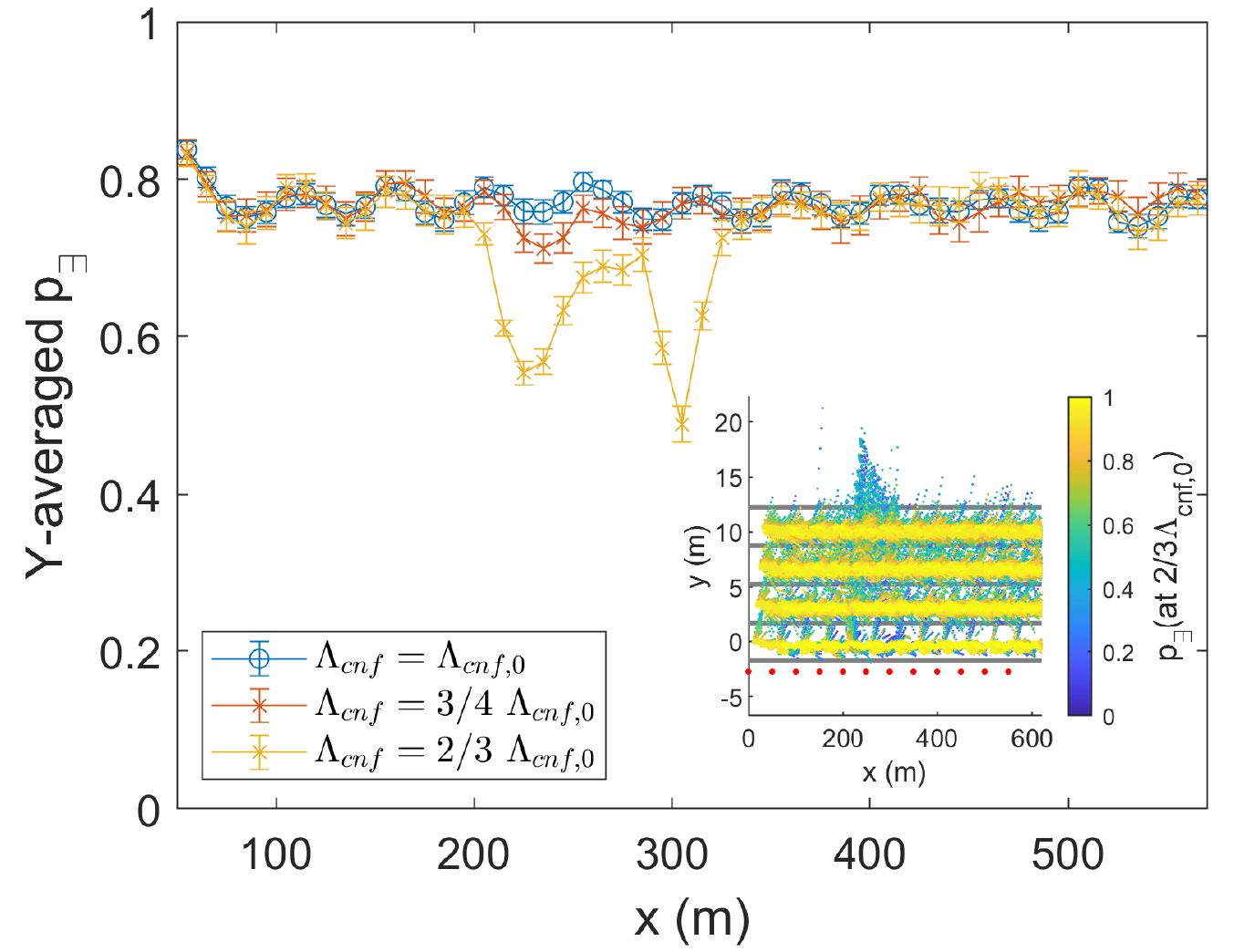}
		\end{subfigure}%
		
\caption{Plausibility signatures of the highway scenario with injected tracking parametrization fault (reduced track confirmation threshold $\thconf$ at sensor $5$). This fault can be identified from the locally reduced average $\pp$ as well as the increased MR of the neighboring sensors.}
\label{fig:trError}
\end{figure*}

This scenario of highway surveillance by a network of Radar sensors is guided by long-term observations within the KoRA9 project \cite{Geissler2018, BMVI2017}. In our setup, a four-lane road segment of about $600m$ is monitored by $12$ Radars at a separation of $50m$, where we discard the first and last sensor for the statistics to minimize finite size effects, see Fig.~\ref{fig:setup_simul1}. The simulation is done in Matlab{\textregistered} \cite{Matlab2019} using the \textit{radarDetectionGenerator} class and decentralized MOT with extended Kalman filters (EKF). Key characteristics are $\fv=(90m, 30^{\circ}, 8^{\circ})$, detection probability $\pd=0.9$, false alarm rate of $\pfa=10^{-6}$, and a track confirmation threshold \cite{Blackman1999} of $\thconf=1.5\ \log(\pd/\pfa)$. To reflect experiences with realistic Radar signal behavior, we implement an additional signal generation up to a range of $100m$, with a reduced likelihood, which results in unexpected observations. The adopted highway traffic model features various vehicle classes such as cars, trucks, buses, and generates speed, lane, and type distributions based on realistic highway statistics \cite{ABDSB2018}.
Vehicle maneuvers such as overtaking are neglected here since they do not impact the observation quality significantly, see also \cite{Geissler2018}.

Fig.~\ref{fig:misError} illustrates the plausibility signatures of an injected miscalibration fault at one of the central sensors (here sensor number 5 located at $x=200m$, see Fig.~\ref{fig:setup_simul1}), representing fault class 1 in Tab.~\ref{tab:type_errors}. The affected Radar is misoriented counter-clockwise by an azimuthal disturbance angle $\Delta \phi$. All objects observed by the affected sensor then appear rotated (in the opposite direction) to the fusion module, which is not aware of the altered sensor pose. As a result, object tracks do not run parallel to the lanes anymore, and coast out of the road boundaries. 
We find that the lane-averaged existence probability $\pp$ is significantly reduced in a locally confined region of $200m \lesssim x \lesssim 300m$, which clearly indicates a dependability issue of nearby sensors (see Fig.~\ref{fig:misError}). The deviation appears on top of undulations in $\pp$ caused by the varying distance of a passing objects to the respective closest sensor, which can be observed also in the absence of faults. 
The inconsistencies of observations made by the faulty sensor with its neighbor sensors leaves a trace in the misses and unexpected object statistics as well: The MR of sensor $5$ is significantly increased, as this sensor will no longer observe all targets close to the lower road boundary, in contradiction to its neighbor sensor $6$. At the same time the UOR is reduced, which can be explained the following way: The dominant contribution of unexpected observations in our model comes from objects that are in the sensing cone but beyond the regular sensing range. As sensor $5$ is misoriented counter-clockwise, this sensor field segment is no longer covering the road, and thus no objects are to appear there.

\begin{figure*}[tbp]
    \centering
		\begin{subfigure}{0.33\textwidth}
			\centering
			\includegraphics[width=1\textwidth]{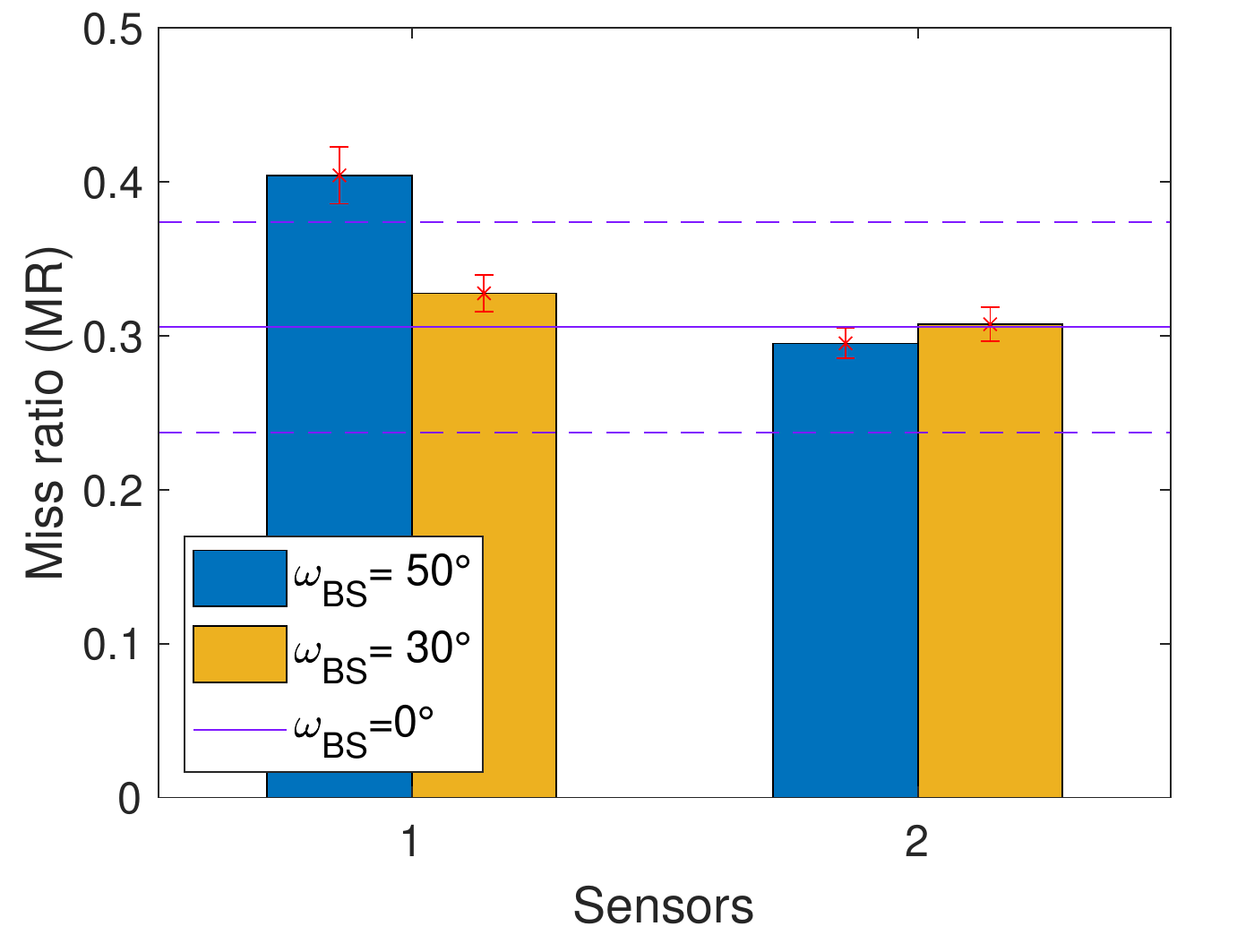}
		\end{subfigure}%
		\begin{subfigure}{0.33\textwidth}
			\centering
			\includegraphics[width=1\textwidth]{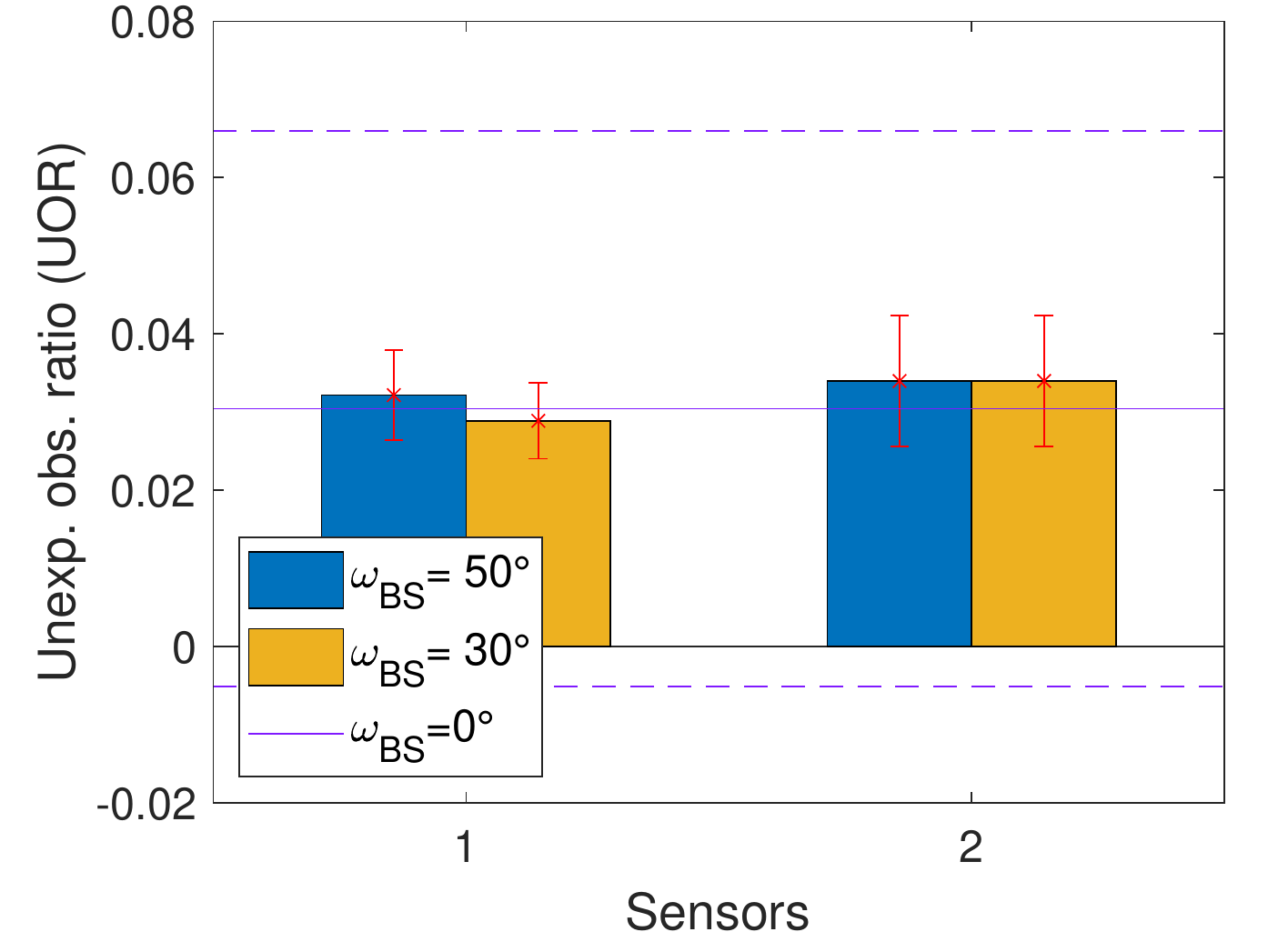}
		\end{subfigure}%
		\begin{subfigure}{0.33\textwidth}
			\centering
			\includegraphics[width=1\textwidth]{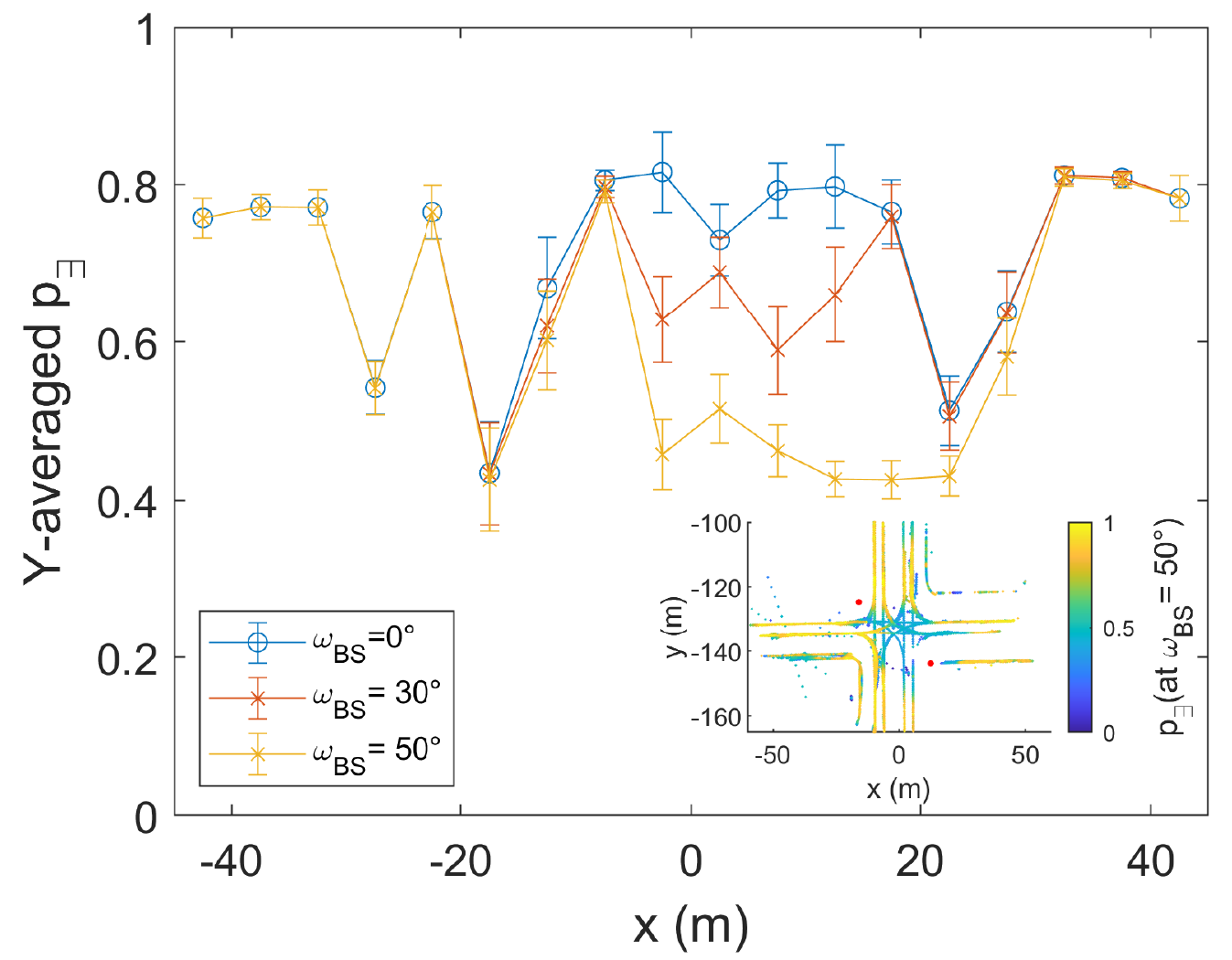}
		\end{subfigure}%
		
\caption{Plausibility signatures derived from the two Lidar sensors covering the intersection shown in Fig.~\ref{fig:CarlaIntersection}. The perception of sensor 1 is impaired by a blind spot of 	azimuthal extension $\wbs$, inducing a high false negative rate.}
\label{fig:dirtError}
\end{figure*}

In the next scenario analyzed in Fig.~\ref{fig:trError} we simulate a fault of class 2 (see Tab.~\ref{tab:type_errors}) by altering the tracking parameters of the test sensor $5$.
While an appropriately parametrized tracking system is quite robust against noise, the tuning of track management parameters has a significant impact on the false object observation rate. A poor parametrization can be caused for example by human intervention, software updates, or -- if machine learning is involved -- by the deployment of non-optimal training sets etc.
Such faults seem a valid concern given that the use of deep learning methods for MOT is a popular trend not only for video applications \cite{Ciaparrone2020}.
To demonstrate tracking errors we here modify  the track confirmation threshold $\thconf$. When reduced with respect to its original value $\thconfo$, tentative tracks are accepted after fewer matching measurement updates, such that more false positive confirmed tracks are generated across the sensor's FoV.
As a consequence, our simulation showcases that the MR of the \textit{neighboring} sensors $4$ and $6$ are significantly increased, since those overlapping sensors can not confirm the large number of false positive tracks generated by sensor 5. On the other hand, this mismatch is effectively compensated by the likewise increased number of overall observations for sensor $5$ itself. In the UOR statistics, no significant fingerprint can be found since the false positive tracks are mostly born in the regular FoV and thus will not be registered as unexpected. Finally, the lane-averaged $\pp$ again gives a clear indication of the locally confined dependability issue. The dip in the projected existence probability here follows a mexican hat potential, which reflects the geometry of the sensor field overlaps.

\begin{figure}[tbp]
	\center
	\includegraphics[width=0.49\textwidth]{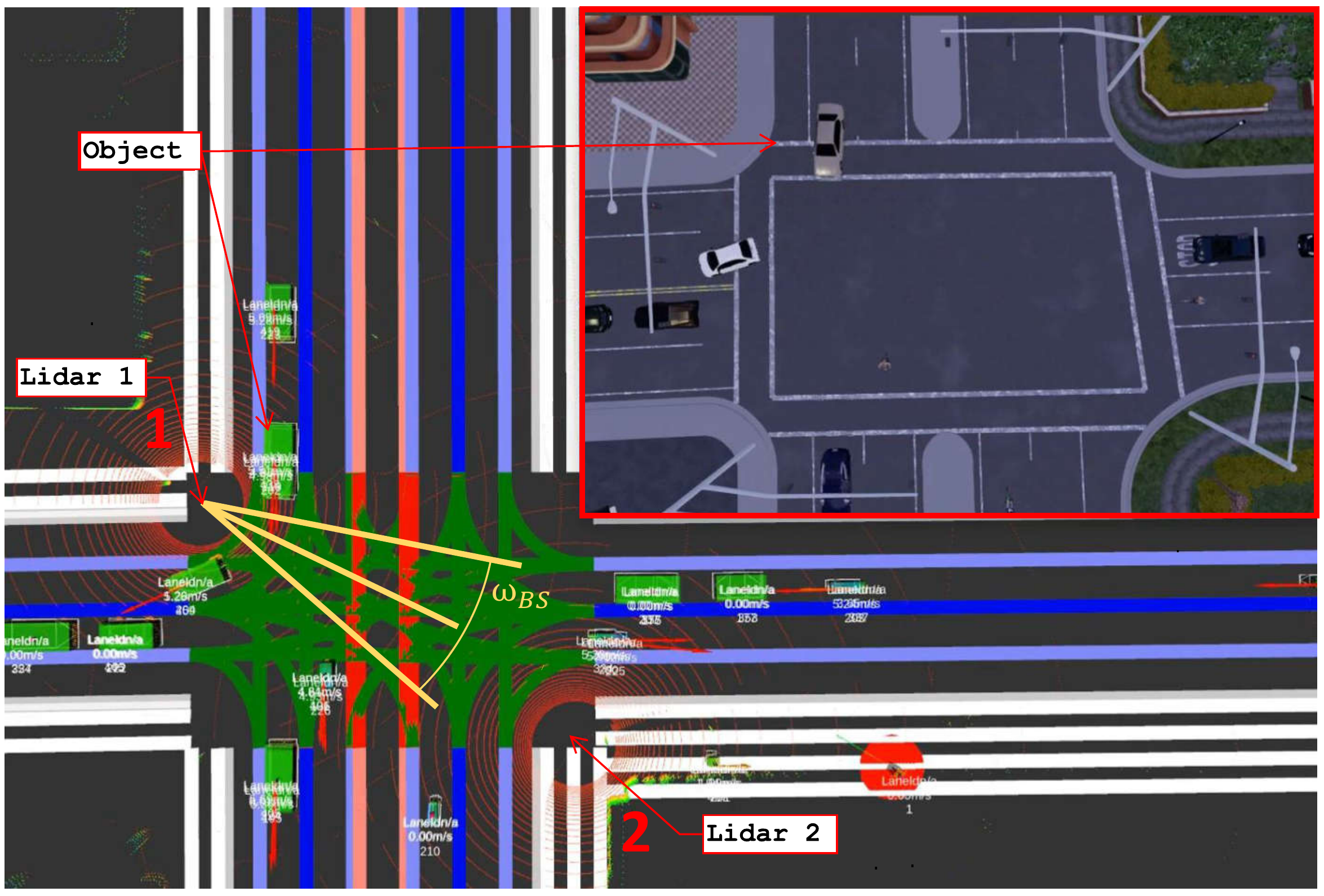}%
	\caption{Overview of the CARLA multi-lane intersection used for the urban scenario. Two Lidar sensors are located on two opposite corners of the junction, cocentric circles indicate the point clouds generated by each Lidar. A blind spot fault with tunable occlusion angle $\wbs$ is injected at sensor 1.}
	\label{fig:CarlaIntersection}
\end{figure}

\subsection{Scenario 2: Urban intersection with Lidar}

According to recent surveys \cite{NHTSA2019}, about $37\%$ of fatal vehicle crashes in the USA are intersection-related. Junctions as critical hotspots will thus benefit in particular from infrastructure assistance, and represent an interesting use case for plausibility-related fault evaluation. In this section, we study an urban intersection scenario that is guided by the real infrastructure testbed in \cite{Fleck2019} (with the difference that Lidars are deployed instead of cameras for ease of the simulation pipeline).
The recordings were performed in the open-source simulator for autonomous driving research CARLA \cite{Dosovitskiy2017}. Based on the Unreal Engine for the rendering, it provides with a set of towns and maps in addition to various sensor models such as RGB cameras, depth sensors, and Lidar sensors.
For this scenario, a four-way intersection of the \textit{Town03} map that resembles the setup in \cite{Fleck2019} was chosen, see Fig.~\ref{fig:CarlaIntersection}. Multiple lanes at each intersection leg, and a large variety of randomly spawned vehicles allow for a diverse urban traffic scenario featuring cars, trucks, cyclists as well as pedestrians roaming on the sidewalks. The recorded scenario has an overall length of $370s$, which is split into quasi-independent intervals of $5s$ each, and on average $18$ tracks are detected per frame.

At this intersection, two Lidar sensors are placed at two opposing corners, with a $\fv = (80m, 360^{\circ}, 28^{\circ})$ that allows to monitor all events taking place at the junction. This layout represents a minimal example setup for accurate object detection, under mitigation of occlusion events.
Each $16$-channel Lidar sensor provides with a cloud of $230400$ points per second, at an update rate of $20\text{Hz}$. 
The point cloud feeds into a deep neural network performing 3D object detection with the PointPillars algorithm, see Refs.~\cite{Geiger2018, Lang2019} for performance and accuracy benchmarks on the KITTI data set. A separate study with synthetic CARLA data was conducted to verify that PointPillars achieves a mean average precision similar to KITTI on the classes car, cyclist and pedestrian, both from a bird-eye-view and a 3D perspective \cite{IntelCarla2020}.
The generated object list includes information about 3D dimension, position, and heading. Subsequently, the extracted bounding boxes are tracked with an EKF. 
CARLA further provides with an autonomous agent behavior: Vehicles stay in their lanes, follow traffic lights and take random turns at intersections (whenever possible), drive at a predefined speed, and do not perform any maneuvers such as overtaking. Another limitation of the CARLA simulator is the fact that the traffic lights at an intersection follow a round-robin schedule, where only one road has a green light at a time. Even though the traffic model in the simulator is simplistic, there are enough actors to produce occlusions.

In Fig.~\ref{fig:dirtError}, we see that the established baseline of the averaged $\pp$ shows a unsteady behavior at the intersection interfaces, even in the absence of injected faults. This can partially be attributed to the fact that vehicles are frequently turning and accelerating in this zones, which poses severe challenges for the trackers.   
A class-3-fault (see Tab.~\ref{tab:type_errors}) is now injected into the simulation, in the form of a blind spot for sensor 1 that represents e.g. a significant mud splash on the Lidar cover. Sensor 1 is now blind to detections under certain azimuthal angles facing the intersection center, where the extension of the blind spot is controlled by the parameter $\wbs$, see Fig.~\ref{fig:CarlaIntersection}.   
The simulation results in Fig.~\ref{fig:dirtError} demonstrate that the observed probability of existence is statistically reduced in the inner junction, given a sufficiently drastic sensor pollution. At the same time, the miss ratio of the affected sensor 1 is increased due to the inconsistent observations with respect to sensor 2. 
Note that this gives us a handle to identify the faulty Lidar within the described plausibilization scheme, in a busy urban environment, even though in this experiment only two overlapping sensors were used.

\section{Summary and conclusion}
\label{sec:summary}

This article investigated whether specific plausibility metrics -- in our model represented by sensor miss ratios, unexpected observation ratios, and the collaborative existence probability derived from various plausibility checks across the monitored region -- can act as fault indicators in complex decentralized perception systems. 
For two realistic roadside infrastructure scenarios, we have tested selected perception faults in simulation, and demonstrated that the resulting statistical fingerprints in the plausibility metrics allow for a conclusive fault detection and evaluation. The injected faults were hereby identified from experience with the real-world smart infrastructure systems KoRA9 and P++.
Tab.~\ref{tab:summary} summarizes the results of this analysis. 

Since the deployed plausibility checks are independent of the input sensor modality, or the underlying perception algorithms, our analysis can be used in particular to verify the integrity of systems with black box behavior. An extension with checks focused on machine learning-based perception faults is envisioned for further research.
The plausibility signatures are the more pronounced the more sensor overlap can be generated. With this work, we therefore emphasize as well the importance of redundant, independent sources of information for fault tolerant automated driving.

\begin{table}[htbp]
\centering
\begin{tabularx}{0.5\textwidth}{X  X} 
			\toprule
			Fault & Characteristic signatures\\ \midrule
			Misoriented sensor pose & Reduced local $\pp$, increased MR for affected sensor, reduced UOR for affected sensor \\ 
			Incorrect parametrization of algorithmic component (tracker threshold) & Reduced local $\pp$, increased MR for other overlapping sensors \\ 
			Sensor polluted  & Reduced local $\pp$, increased MR for affected sensor \\ \bottomrule
\end{tabularx}
\vspace{0.3cm}
\caption{Overview of the simulation results obtained in Sec.~\ref{sec:results}, confer also with Tab.~\ref{tab:type_errors}. The three selected empirical perception faults analyzed in the two smart infrastructure scenarios lead to distinct fingerprints in the plausibility metrics.}
\label{tab:summary}
\end{table}




\bibliographystyle{splncs} 
\bibliography{PlausBib}

\end{document}